\documentclass[aps,prd,twocolumn,linenumbers,superscriptaddress,noshowpacs,floatfix]{revtex4}
\usepackage{graphicx}
\usepackage{dcolumn}
\usepackage{bm}
\usepackage{xcolor}
\usepackage{lineno}

\usepackage{subfigure}
\usepackage{amsmath,epsfig}
\usepackage{epstopdf}
\usepackage{multirow}
\usepackage{hhline}
\usepackage{afterpage}
\usepackage{float}
\usepackage{hyperref} 
\usepackage{booktabs}

\begin{document}

\title{A new approach to dark photon}

\author{Jie Tang}

\email{tangj@seu.edu.cn}

\author{Pei-Hong Gu}

\email{phgu@seu.edu.cn}

\affiliation{School of Physics, Jiulonghu Campus, Southeast University, Nanjing 211189, China}

\begin{abstract}

Over the past few decades, the hypothetically dark photon has been extensively studied from both phenomenological and experimental perspectives. It should be noted that the local symmetry for dark photon does not gauge the standard model Higgs scalar and chiral fermions. In this paper, we show that an artificially introduced $U(1)_X^{}$ gauge group for dark photon and the standard model $U(1)_Y^{}$ gauge group for hypercharge can be simultaneously born from two $U(1)_1^{}\times U(1)_2^{}$ gauge groups under which the standard model scalar and fermions carry the same $U(1)_1^{}$ and $U(1)_2^{}$ charges without causing any gauge anomalies. We further introduce a spontaneously broken mirror symmetry between the $U(1)_1^{}$ and $U(1)_2^{}$ gauge groups so that the $U(1)_1^{}$ and $U(1)_2^{}$ gauge couplings can acquire a small difference at one-loop level and hence the $U_X^{} \times U_Y^{}$ kinetic mixing can be highly suppressed in a natural way.

\end{abstract}


\maketitle

\section{Introduction}

The standard model (SM) of particle physics is a gauge theory with the $SU(3)_c^{}\times SU(2)_L^{} \times U(1)_Y^{}$ local symmetries. Usually the SM Higgs scalar and chiral fermions can not be overall gauged by additionally local symmetries in the absence of proper non-SM fermions. Otherwise, the theory will suffer a disaster of gauge anomaly unless the additional symmetry is a new $U(1)$ gauge group and its associated quantum numbers are proportional to the existing $U(1)$ quantum numbers. However, the existing $U(1)$ gauge group can no longer be the SM $U(1)_Y^{}$ gauge group in this situation. Actually the physical $U(1)_Y^{}$ gauge group now should be a linear combination of the new $U(1)$ gauge group and the existing $U(1)$ gauge group. 

On the other hand, the dark photon \cite{okun1982,holdom1986,fh1991} from an artificially introduced $U(1)_X^{}$ gauge group is a hypothetical particle with very attractive phenomena and has been  extensively studied by many theorists and experimentalists \cite{fgl2020,cmohv2021,abudinen2023,fgkt2017,abreu2023,alekhin2016,ght2016,bgst2018,ccl2017,alpigiani2019,rp2009,fppr2014,bjw2016,ppru2018,ikns2020,mdw2020,cfmw2020,lfg2020,clmspr2023,phj2024,ganliu2023,dczjw2025,agbbkpskss2024,mcphhjr2024,hzhou2024,tnxu2025,tbclcv2025,dczw2025,gllm2025,hhs2025,agms2025,cpy2025,mtk2025,polesello2025,bck2025,eanb2025,xysz2025,xuzhou2025,hegdk2025,alqv2026,ybyh2026,edsmv2026,rsysb2026,prs2026}. The interactions of the dark photon to the SM are assumed from an abelian kinetic mixing between the SM $U(1)_Y^{}$ gauge group and the new $U(1)_X^{}$ gauge group. This $U(1)_X^{}\times U(1)_Y^{}$ kinetic mixing can not be sizable due to the stringent constraints from different experiments such as low energy colliders, meson decays, beam dump experiments, high-energy colliders and so on. However, the smallness of a renormalisable $U(1)$ kinetic mixing can not be taken for granted.

As the local symmetry for dark photon does not gauge the SM Higgs scalar and chiral fermions, the unconfirmed dark photon appears to warrant a compelling motivation from a theoretical perspective. In this paper, we show that an artificially introduced $U(1)_X^{}$ gauge group for dark photon and the SM $U(1)_Y^{}$ gauge group for hypercharge can be simultaneously induced by two $U(1)_1^{}\times U(1)_2^{}$ gauge groups under which the SM Higgs scalar and chiral fermions carry the same $U(1)_1^{}$ and $U(1)_2^{}$ charges without causing any gauge anomalies. Remarkably, the $U(1)_X^{}\times U(1)_Y^{}$ kinetic mixing would not emerge at all if the $U(1)_1^{}$ gauge coupling was exactly equal to the $U(1)_2^{}$ gauge coupling. We further introduce a spontaneously broken mirror symmetry between the $U(1)_1^{}$ and $U(1)_2^{}$ gauge groups so that the $U(1)_1^{}$ and $U(1)_2^{}$ gauge couplings can acquire a small difference at one-loop level and hence the $U_X^{} \times U_Y^{}$ kinetic mixing can be highly suppressed in a natural way.

\section{Key thought}

The basic model contains one Higgs scalar and three generations of chiral fermions, i.e.
\begin{eqnarray}
\label{smfields}
&&\phi\!\left(\!\begin{array}{l}1,2,-\frac{1}{2},-\frac{1}{2}\end{array}\!\right) \!= \!\left[\begin{array}{l}\phi^{0}_{}\\
[2mm]
\phi^{-}_{}\end{array}\right];~~q_{Li}^{}\!\left(\!\begin{array}{l}3,2,+\frac{1}{6},+\frac{1}{6}\end{array}\!\right)\!=\!\left[\begin{array}{l}u^{}_{Li}\\
[2mm]
d^{}_{Li}\end{array}\right],\nonumber\\
[3mm]
&&d_{Ri}^{}\!\left(\!\begin{array}{l}3,1,-\frac{1}{3},-\frac{1}{3}\end{array}\!\right),~~u_{Ri}^{}\!\left(\!\begin{array}{l}3,1,+\frac{2}{3},+\frac{2}{3}\end{array}\!\right),\nonumber\\
[3mm]
&&l_{Li}^{}\!\left(\!\begin{array}{l}1,2,-\frac{1}{2},-\frac{1}{2}\end{array}\!\right)\!=\!\left[\begin{array}{l}\nu^{}_{Li}\\
[2mm]
e^{}_{Li}\end{array}\right],~~e_{Ri}^{}\!\left(\!\begin{array}{l}1,1,-1,-1\end{array}\!\right),\nonumber\\
[2mm]
&&(i=1,2,3)\,.
\end{eqnarray}
Here and thereafter the brackets following the fields describe the transformations under the $SU(3)_c^{}\times SU(2)_L^{} \times U(1)_1^{} \times U(1)_2^{}$ gauge groups. Clearly, these fermions do not cause any gauge anomalies. We will clarify later the scalar $\phi$ is the SM Higgs scalar, the fermions $d_{Li,Ri}^{}$, $u_{Li,Ri}^{}$ and $e_{Li,Ri}^{}$ are the SM charged fermions while the fermions $\nu_{Li}^{}$ are the SM neutral neutrinos.

For simplicity, we do not write down the full kinetic terms. Instead, we only give the parts involving the $ U(1)_1^{}$ and $U(1)_2^{}$ gauge groups, i.e.
\begin{eqnarray}
\label{smkinetic}
\mathcal{L}_K^{}&\supset& -\frac{1}{4} X_{1\mu\nu}^{} X_{1} ^{\mu\nu} - \frac{1}{4} X_{2\mu\nu}^{} X_{2}^{\mu\nu} - \frac{\epsilon_{12}^{}}{2} X_{1\mu\nu}^{} X_{2}^{\mu\nu} \nonumber\\
[2mm]
&&+\left(D_\mu^{} \phi\right)^\dagger_{}D^\mu_{} \phi + \sum_{i}^{}\left(i \bar{q}_{Li}^{}\gamma^\mu_{} D_\mu^{} q_{Li}^{}  \right.\nonumber\\
[2mm]
&&+ i \bar{d}_{Ri}^{}\gamma^\mu_{} D_\mu^{} d_{Ri}^{}+ i \bar{u}_{Ri}^{}\gamma^\mu_{} D_\mu^{} u_{Ri}^{} + i \bar{l}_{Li}^{}\gamma^\mu_{} D_\mu^{} l_{Li}^{}\nonumber\\
[2mm]
&&\left.+ i \bar{e}_{Ri}^{}\gamma^\mu_{} D_\mu^{} e_{Ri}^{} \right)\,.
\end{eqnarray}
Here the gauge field strength tensors are 
\begin{eqnarray}
\label{smfieldstength}
X_{1\mu\nu}^{}= \partial_\mu^{} X_{1\nu}^{} - \partial_\nu^{} X_{1\mu}^{}\,,~~X_{2\mu\nu}^{}= \partial_\mu^{} X_{2\nu}^{} - \partial_\nu^{} X_{2\mu}^{}\,,~~
\end{eqnarray}
while the covariant derivatives include
\begin{eqnarray}
\label{covariant}
D_\mu^{} \phi &\supset& \left(\partial_\mu^{}- i  \frac{1}{2}  g_1^{}X^{}_{1\mu}- i  \frac{1}{2}  g_2^{}X^{}_{2\mu}\right) \phi\,,\nonumber\\
[2mm]
D_\mu^{} q_{Li}^{}&\supset& \left(\partial_\mu^{}+ i  \frac{1}{6}  g_1^{}X^{}_{1\mu}+ i  \frac{1}{6}  g_2^{}X^{}_{2\mu} \right) q_{Li}^{}\,,\nonumber\\
[2mm]
D_\mu^{} d_{Ri}^{}&\supset& \left(\partial_\mu^{}- i  \frac{1}{3}  g_1^{}X^{}_{1\mu}- i  \frac{1}{3}  g_2^{}X^{}_{2\mu}\right) d_{Ri}^{}\,,\nonumber\\
[2mm]
D_\mu^{} u_{Ri}^{}&\supset&  \left(\partial_\mu^{}+ i  \frac{2}{3}  g_1^{}X^{}_{1\mu}+ i  \frac{2}{3}  g_2^{}X^{}_{2\mu}\right) u_{Ri}^{}\,,\nonumber\\
[2mm]
D_\mu^{} L_{Li}^{}&\supset& \left(\partial_\mu^{} - i \frac{1}{2}  g_1^{} X^{}_{1\mu} - i  \frac{1}{2}  g_2^{}X^{}_{2\mu}\right)l_{Li}^{}\,,\nonumber\\
[2mm]
D_\mu^{} e_{Ri}^{}&\supset& \left(\partial_\mu^{}- i g_1^{} X^{}_{1\mu} - i  g_2^{}X^{}_{2\mu}\right)e_{Ri}^{}\,,
\end{eqnarray}
with $g_{1,2}^{}$ being the $U(1)_{1,2}^{}$ gauge couplings. Moreover, the $[U(1)_1^{}\times U(1)_2^{}]$-kinetic-mixing constant $\epsilon_{12}^{}$ is a physical parameter which can not be scaled away.

The Higgs scalar and the chiral fermions (\ref{smfields}) also have the following Yukawa couplings,
\begin{eqnarray}
\mathcal{L}_Y^{}&=&-\sum_{ij}^{}\left(y_{d}^{ij}\bar{q}_{Li}^{}\tilde{\phi} d_{Rj}^{}+ y_{u}^{ij}\bar{q}_{Li}^{}\phi u_{Rj}^{} + y_{e}^{ij}\bar{l}_{Li}^{}\tilde{\phi} e_{Rj}^{}\right.\nonumber\\
[2mm]
&&\left.+\textrm{H.c.}\right)~~\textrm{with}~~\tilde{\phi}=i\tau_2^{} \phi^\ast_{}\,,
\end{eqnarray}
the same as those in the SM. After the Higgs scalar $\phi$ develops its vacuum expectation value, the charged fermions $d_{Li,Ri}^{}$, $u_{Li,Ri}^{}$, and $e_{Li,Ri}^{}$ can obtain their Dirac mass terms. As for the neutral neutrinos $\nu_{Li}^{}$, they still keep massless. However, the conventional ideas for generating the neutrino masses like the famous seesaw mechanism can be directly adopted under the present symmetries. The details are not shown again. 

\section{ Dark photon}

From the covariant derivatives (\ref{covariant}), we conveniently rotate the $U(1)_{1,2}^{}$ gauge fields $X_{1\mu,2\mu}^{}$ to be
\begin{eqnarray}
X'^{}_{1\mu} =\frac{g_1^{} X_{1\mu}^{} + g_2^{}X_{2\mu}^{}}{\sqrt{g_1^2 +g_2^2} }  \,,~~X'^{}_{2\mu} =\frac{ g_1^{} X_{2\mu}^{} - g_2^{}X_{1\mu}^{}  }{\sqrt{g_1^{2} + g_2^2}}\,,\end{eqnarray}
and then rewrite their kinetic terms to be 
\begin{eqnarray}
\mathcal{L}_K^{}&\supset&-\frac{1}{4} X_{1\mu\nu}^{} X_{1} ^{\mu\nu} - \frac{1}{4} X_{2\mu\nu}^{} X_{2}^{\mu\nu} - \frac{\epsilon_{12}^{}}{2} X_{1\mu\nu}^{} X_{2}^{\mu\nu} \nonumber\\
[2mm]
&=& -\frac{1}{4} \left(1+\frac{2 \epsilon_{12}^{} g_1^{} g_2^{}}{g_1^2 +g_2^2} \right)X'^{}_{1\mu\nu} X'^{\mu\nu}_{1} \nonumber\\
[2mm]
&&- \frac{1}{4} \left(1-\frac{2 \epsilon_{12}^{} g_1^{} g_2^{}}{g_1^2 +g_2^2} \right) X'^{}_{2\mu\nu} X'^{\mu\nu}_{2}\nonumber\\
[2mm]
&& - \frac{\epsilon_{12}^{} \left(g_1^2 -g_2^2 \right)}{2\left(g_1^2 +g_2^2\right)} X'^{}_{1\mu\nu} X'^{\mu\nu}_{2} \,.
\end{eqnarray}
Therefore, we eventually obtain the SM $U(1)_Y^{}$ gauge field $B_\mu^{}$ for hypercharge and the new $U(1)_X^{}$ gauge field $X_\mu^{}$ for dark photon, i.e.
\begin{eqnarray}
\mathcal{L}_K^{}&\supset&-\frac{1}{4} B_{\mu\nu}^{} B_{} ^{\mu\nu} - \frac{1}{4} X_{\mu\nu}^{} X_{}^{\mu\nu} - \frac{\epsilon}{2} B_{\mu\nu}^{} X_{}^{\mu\nu} ~~\textrm{with}\nonumber\\
[2mm]
\label{yfield}
&&B^{}_{\mu} =\sqrt{1+\frac{2 \epsilon_{12}^{} g_1^{} g_2^{}}{g_1^2 +g_2^2} } \frac{g_1^{} X_{1\mu}^{} + g_2^{}X_{2\mu}^{}}{\sqrt{g_1^2 +g_2^2} }  \,,\\
[2mm]
\label{xfield}
&&X^{}_{\mu} =\sqrt{1-\frac{2 \epsilon_{12}^{} g_1^{} g_2^{}}{g_1^2 +g_2^2} } \frac{ g_1^{} X_{2\mu}^{} - g_2^{}X_{1\mu}^{}  }{\sqrt{g_1^{2} + g_2^2}} \,,\\
[2mm]
\label{kinetic}
&&~~~\epsilon=  \frac{\epsilon_{12}^{} \left(g_1^2 -g_2^2 \right)}{\left(g_1^2 +g_2^2\right)}  \left/ \sqrt{1-\frac{4 \epsilon_{12}^{2} g_1^{2} g_2^{2}}{\left(g_1^2 +g_2^2\right)^2_{}} }\right.\,.\end{eqnarray}
The above kinetic terms can be diagonalised by making a non-unitary transformation \cite{fh1991}. The kinetic-mixing parameter (\ref{kinetic}) thus should be constrained in the range $|\epsilon|<1$ for the requirement of positive kinetic energy.

It is easy to see that the Higgs scalar and chiral fermions (\ref{smfields}) definitely can be identified to the SM ones because their covariant derivatives (\ref{covariant}) now have become the SM appearance, i.e. 
\begin{eqnarray}
\label{smcovariant}
&&D_\mu^{} \phi \supset - i  \frac{1}{2}  g' B^{}_{\mu}  \phi\,,~~
D_\mu^{} q_{Li}^{}\supset + i  \frac{1}{6}  g' B^{}_{\mu} q_{Li}^{}\,,\nonumber\\
[2mm]
&&D_\mu^{} d_{Ri}^{}\supset - i  \frac{1}{3} g' B^{}_{\mu} d_{Ri}^{}\,,~~D_\mu^{} u_{Ri}^{}\supset  + i  \frac{2}{3} g' B^{}_{\mu} u_{Ri}^{}\,,\nonumber\\
[2mm]
&& D_\mu^{} L_{Li}^{}\supset - i \frac{1}{2} g' B^{}_{\mu}  l_{Li}^{}\,,~~D_\mu^{} e_{Ri}^{}\supset - i g' B^{}_{\mu} e_{Ri}^{}\,.
\end{eqnarray}
Here the $U(1)_Y^{}$ gauge coupling $g'$ has been determined by
\begin{eqnarray}
\label{gprime}
g'= \sqrt{g_1^2 + g_2^2 }  \left/ \sqrt{1+\frac{2 \epsilon_{12}^{} g_1^{} g_2^{}}{g_1^2 +g_2^2} }\right.  \,.
\end{eqnarray}
Subsequently, the $U(1)_X^{}$ gauge coupling $g_X^{}$ is given by 
\begin{eqnarray}
\label{gx}
g_X^{} &=& \sqrt{g_1^2 + g_2^2 }  \left/ \sqrt{1-\frac{2 \epsilon_{12}^{} g_1^{} g_2^{}}{g_1^2 +g_2^2} }\right.  \nonumber\\
[2mm]
&=& g'  \sqrt{1+\frac{2 \epsilon_{12}^{} g_1^{} g_2^{}}{g_1^2 +g_2^2} }  \left/ \sqrt{1-\frac{2 \epsilon_{12}^{} g_1^{} g_2^{}}{g_1^2 +g_2^2} }\right.\,.
\end{eqnarray}

So far the $U(1)_X^{}$ gauge symmetry has not been spontaneously broken. Consequently, the $U(1)_X^{}$ gauge field $X_\mu^{}$ can only contribute a massless dark photon. If we expect a massive dark photon, we should consider a proper Higgs scalar $\xi$ to spontaneously break the $U(1)_{1}^{}\times U(1)_{2}^{}$ gauge symmetries down to the $U(1)_Y^{}$ gauge symmetry, i.e.
\begin{eqnarray}
\label{breaking}
U(1)_1^{}\times U(1)_2^{} \stackrel{\xi=\frac{1}{\sqrt{2}}\left(v_\xi^{} + h_\xi^{} \right)\exp\left(iG_\xi^{}/v_\xi^{}\right)}{-\!\!\!-\!\!\!-\!\!\!-\!\!\!-\!\!\!-\!\!\!-\!\!\!-\!\!\!-\!\!\!-\!\!\!-\!\!\!-\!\!\!-\!\!\!-\!\!\!-\!\!\!-\!\!\!-\!\!\!-\!\!\!-\!\!\!-\!\!\!\rightarrow } U(1)_Y^{}\,,
\end{eqnarray}
where $v_\xi^{}$, $h_\xi^{}$ and $G_\xi^{}$ are the vacuum expectation value, the Higgs boson and the would-be Goldstone boson, respectively. For this purpose, the Higgs scalar $\xi$ can only carry the quantum numbers as below,
\begin{eqnarray}
\label{xhiggs}
\xi\left(1,1,+\frac{g_2^{}}{g_1^{}} x,-\frac{g_1^{}}{g_2^{}} x \right)\,,
\end{eqnarray}
with $x$ being a nonzero constant. Obviously we indeed introduce a $U(1)_X^{}$ Higgs scalar. Accordingly, the $U(1)_X^{}$ gauge field $X_\mu^{}$ can eat the would-be Goldstone boson $G_\xi^{}$. The dark photon $X_\mu^{}$ then can acquire the following mass term,
\begin{eqnarray}
\label{dpmass}
\mathcal{L}&\supset& \frac{1}{2} m_X^2 X_{\mu}^{} X^{\mu}_{}~~\textrm{with} ~~m_X^2 = x^2_{} g_X^2 v_\xi^2 \,.
\end{eqnarray}

Note for the known $U(1)_Y^{}$ gauge coupling $g'$ in Eq. (\ref{gprime}), the new $U(1)_X^{}$ gauge coupling $g_X^{}$ in Eq. (\ref{gx}) can not become very small unless there is a fine tune among the $[U(1)_1^{} \times U(1)_2^{}]$-kinetic-mixing parameter $\epsilon_{12}^{}$ and the $U(1)_{1,2}^{}$ gauge couplings $g_{1,2}^{}$. Although the vacuum expectation value $v_\xi^{}$ usually does not spontaneously develop an extremely small vacuum expectation value, the dark photon $X_{\mu}^{}$ can still acquire an ultralight mass $m_X^{}$ in Eq. (\ref{dpmass}). This is because the $[U(1)_1^{}\times U(1)_2^{}]$-charge constant $x$ can be assumed very small. Under this circumstance, we can be allowed to realize a tiny $[U(1)_X^{} \times U(1)_Y^{}]$-kinetic-mixing parameter $\epsilon$ in Eq. (\ref{kinetic}) as long as the $[U(1)_1^{} \times U(1)_2^{}]$-kinetic-mixing parameter $\epsilon_{12}^{}$ has a small value or the $U(1)_{1,2}^{}$ gauge couplings $g_{1,2}^{}$ have a small difference.

\section{Slightly kinetic mixing }

If the dark photon mass $m_X^{}$ is very light, the $[U(1)_X^{}\times U(1)_Y^{}]$-kinetic-mixing parameter $\epsilon$ should be very small to match the null results from dark photon search experiments. In order to naturally explain such a slight mixing, we shall consider a novel scheme as follows.

The first step is to introduce a mirror symmetry between the $U(1)_1^{}$ and $U(1)_2^{}$ gauge groups, i.e.
\begin{eqnarray}
\label{mirror}
 U(1)_{1}^{}\stackrel{M_{12}^{}}{\leftarrow\!\!\!-\!\!\!-\!\!\!-\!\!\!\rightarrow} U(1)_{2}^{} &:& ~~X_{1\mu}^{} \stackrel{M_{12}^{}}{\leftarrow\!\!\!-\!\!\!-\!\!\!-\!\!\!\rightarrow} X_{2\mu}^{}\nonumber\\
 [2mm]
 &&~~\textrm{with}~~g_1^{}=g_2^{} \equiv  g_{12}^{}\,.
 \end{eqnarray}
The SM Higgs scalar and chiral fermions (\ref{smfields}) as well as the $U(1)_X^{}$ Higgs scalar (\ref{xhiggs}) then respect this mirror symmetry according to the following rule,
\begin{eqnarray}
 \textrm{SM} \stackrel{M_{12}^{}}{\leftarrow\!\!\!-\!\!\!-\!\!\!-\!\!\!\rightarrow} \textrm{SM} \,, ~~\xi \stackrel{M_{12}^{}}{\leftarrow\!\!\!-\!\!\!-\!\!\!-\!\!\!\rightarrow} \xi^\ast_{}\,. 
\end{eqnarray}
As a result, the two gauge couplings $g_{1,2}^{}$ should exactly keep equal even if all radiative corrections are taken into account. From Eq. (\ref{kinetic}), we thus can guarantee a vanishing $U(1)_X^{}\times U(1)_Y^{}$ kinetic mixing at this stage.

 \begin{figure*}
\centering
\includegraphics[scale=0.75]{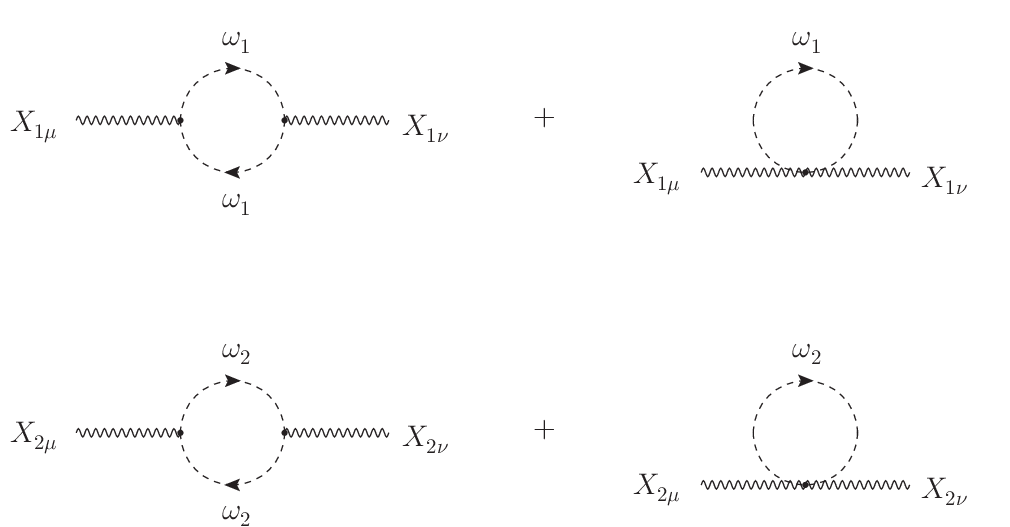} \caption{\label{couplings} The scalar $\omega_1^{}$ carrying the $U(1)_1^{}$ charge and the scalar $\omega_2^{}$ carrying the $U(1)_2^{}$ charge respectively contribute to the kinetic terms of the $U(1)_1^{}$ and $U(1)_2^{}$ gauge fields at one-loop level. The scalars $\omega_1^{}$ and $\omega_2^{}$ have a small mass difference as a result of the spontaneous breaking of a mirror symmetry between the $U(1)_1^{}$ and $U(1)_2^{}$ gauge groups. Consequently there is a small difference between the gauge couplings of the  $U(1)_1^{}$ and $U(1)_2^{}$ gauge groups.}
\end{figure*}

Subsequently we extend the model by the following scalars, 
\begin{eqnarray}
\label{mirrorscalars}
\sigma\!\left(\!\begin{array}{l}1,1,0,0\end{array}\!\right) &\stackrel{M_{12}^{}}{\leftarrow\!\!\!-\!\!\!-\!\!\!-\!\!\!\rightarrow}& -\sigma\!\left(\!\begin{array}{l}1,1,0,0\end{array}\!\right)\,, \nonumber\\
[2mm]
\omega_1^{}\!\left(\!\begin{array}{l}1,1,+8,0\end{array}\!\right)& \stackrel{M_{12}^{}}{\leftarrow\!\!\!-\!\!\!-\!\!\!-\!\!\!\rightarrow}& \omega_2^{}\!\left(\!\begin{array}{l}1,1,0,+8\end{array}\!\right)\,,\nonumber\\
[2mm]
\eta \!\left(\!\begin{array}{l}1,1,-4,+4\end{array}\!\right) &\stackrel{M_{12}^{}}{\leftarrow\!\!\!-\!\!\!-\!\!\!-\!\!\!\rightarrow}& \eta^\ast_{}\!\left(\!\begin{array}{l}1,1,+4,-4\end{array}\!\right)\,, \nonumber\\
[2mm]
\delta \!\left(\!\begin{array}{l}1,1,-2,-2\end{array}\!\right)&\stackrel{M_{12}^{}}{\leftarrow\!\!\!-\!\!\!-\!\!\!-\!\!\!\rightarrow}& \delta\!\left(\!\begin{array}{l}1,1,-2,-2\end{array}\!\right)\,.\end{eqnarray}
For simplicity, we only exhibit the part of the interactions involving the above scalars, i.e. 
\begin{eqnarray}
\label{newinteractions}
\mathcal{L}&\supset& -\mu_{\sigma\omega}^{}\sigma \left(\omega_1^\dagger \omega_1^{} - \omega_2^\dagger \omega_2^{}\right)- \frac{1}{2}\kappa_{\eta\omega\delta}^{} \left[\left(\eta \omega_1^{}\delta^2_{} + \eta^\ast_{} \omega_2^{}\delta^2_{}\right)\right.\nonumber\\
[2mm]
&&\left.+\textrm{H.c.}\right]  -\sum_{ij}^{} \left(\frac{1}{2}y^{ij}_{\delta}\delta \bar{e}_{Ri}^{} e_{Rj}^{c}+\textrm{H.c.}\right)\,.  
\end{eqnarray}
After the real singlet scalar $\sigma$ spontaneously breaks the mirror symmetry $M_{12}^{}$ \cite{bm1989}, the two scalars $\omega_1^{}$ and $\omega_2^{}$ can obtain a mass difference, i.e. 
\begin{eqnarray}
\label{msplit}
m_{\omega_1}^2 - m_{\omega_2}^2 = \mu_{\sigma\omega}^{} v_{\sigma}^{}~~\textrm{for}~~\sigma=v_\sigma^{} + h_\sigma^{}\,,
\end{eqnarray}
with $v_\sigma^{}$ being the vacuum expectation value and $h_\sigma^{}$ being the Higgs boson. It is possible to replace the dilepton scalar $\delta$ by other dilepton scalars, diquark scalars or leptoquark scalars if we arrange the scalars $\omega_{1,2}^{}$ and $\eta$ for proper quantum numbers. For simplicity, we do not show the details of these alternatives.

Since the scalar $\omega_1^{}$ carries the $U(1)_1^{}$ charge rather than the $U(1)_2^{}$ charge while the scalar $\omega_2^{}$ carries the $U(1)_2^{}$ charge rather than the $U(1)_1^{}$ charge, they can respectively contribute to the kinetic terms of the $U(1)_1^{}$ and $U(1)_2^{}$ gauge fields at one-loop level, i.e.
\begin{eqnarray}
\mathcal{L}_K^{}&\supset&-\frac{1}{4} \left(1+c_1^{}\right)X_{1\mu\nu}^{} X_{1} ^{\mu\nu} - \frac{1}{4}\left(1+c_2^{}\right) X_{2\mu\nu}^{} X_{2}^{\mu\nu} \nonumber\\
[2mm]
&&- \frac{\epsilon_{12}^{}}{2} X_{1\mu\nu}^{} X_{2}^{\mu\nu} \,.
\end{eqnarray}
The relevant diagrams are shown in Fig. \ref{couplings}. In the minimal subtraction scheme, the radiative corrections $c_{1,2}^{}$ are calculated by 
\begin{eqnarray}
c_1^{}=\frac{8g_{12}^2 }{3\pi^2_{}} \ln \left( \frac{m_{\omega_1}^2}{\mu_{}^2}\right)\,,~~c_2^{}=\frac{8g_{12}^2 }{3\pi^2_{}} \ln \left( \frac{m_{\omega_2}^2}{\mu_{}^2}\right)\,,
\end{eqnarray}
with $\mu$ being the renormalization scale.

Now the $U(1)_{1,2}^{}$ gauge couplings $g_{1,2}^{}$ and the $[U(1)_1^{}\times U(1)_2^{}$]-kinetic-mixing parameter $\epsilon_{12}^{}$ in the previous demonstrations should be adjusted by 
\begin{eqnarray}
&&g_{1}^{}  \rightarrow \bar{g}_1^{}=\frac{g_{12}^{} }{\sqrt{1+c_1^{}}} \,,~~g_{2}^{} \rightarrow \bar{g}_2^{}=\frac{g_{12}^{} }{\sqrt{1+c_2^{}}}\,,\nonumber\\
[2mm]
&&\epsilon_{12}^{}\rightarrow \bar{\epsilon}_{12}^{}=\frac{\epsilon_{12}^{}}{\sqrt{(1+c_1^{})(1+c_2^{})}}\,.
\end{eqnarray}
In consequence, the $[U(1)_X^{}\times U(1)_Y^{}]$-kinetic-mixing parameter $\epsilon$ in Eq. (\ref{kinetic}) arrives at a nonzero value, i.e. 
\begin{eqnarray}
\epsilon &=&  \frac{\bar{\epsilon}_{12}^{} \left(\bar{g}_1^2 -\bar{g}_2^2 \right)}{\left(\bar{g}_1^2 +\bar{g}_2^2\right)}  \left/ \sqrt{1-\frac{4 \bar{\epsilon}_{12}^{2} \bar{g}_1^{2} \bar{g}_2^{2}}{\left(\bar{g}_1^2 +\bar{g}_2^2\right)^2_{}} }\right.\nonumber\\
[2mm]
&=&\frac{\bar{\epsilon}_{12}^{}\left(c_2^{}-c_1^{}\right)}{2+c_1^{}+c_2^{}}  \left/ \sqrt{1-\frac{4 \bar{\epsilon}_{12}^{2} }{\left(2+c_1^{}+c_2^{}\right)^2_{}} }\right.  \,,
\end{eqnarray}
which can be highly suppressed in the limiting case as below, 
\begin{eqnarray}
\epsilon &\propto &  \ln\left(\frac{m_{\omega_2}^2}{m_{\omega_1}^2}\right)\ll 1~~\textrm{for}~~m_{\omega_1}^2 - m_{\omega_2}^2 =\mu_{\sigma\omega}^{} v_{\sigma}^{} \ll m_{\omega_{1,2}}^2\,.\nonumber\\
[2mm]
&&
\end{eqnarray}

It should be noted that the scalar $\omega_{1}^{}$ indeed carries a $U(1)_Y^{}$ hypercharge $+4$ and a $U(1)_X^{}$ charge $-4$, the scalar $\omega_{2}^{}$ indeed carries a $U(1)_Y^{}$ hypercharge $+4$ and a $U(1)_X^{}$ charge $+4$, the scalar $\delta$ indeed carries a $U(1)_Y^{}$ hypercharge $-2$, while the scalar $\eta$ indeed carries a $U(1)_X^{}$ charge $+4$, according to the definitions (\ref{yfield}-\ref{xfield}) of the $U(1)_X^{}$ and $U(1)_Y^{}$ gauge fields. In the presence of the quartic scalar couplings and the Yukawa couplings in Eq. (\ref{newinteractions}), the scalars $\omega_{1,2}^{}$ and $\delta$ would not leave any relic density due to their fast decays. As for the SM-singlet scalar $\eta$, it can be allowed to act as a stable dark matter. This stability can be guaranteed by imposing an unbroken $Z_2^{}$ discrete symmetry under which only the scalars $\omega_{1,2}$ and $\eta$ take an odd parity. Alternatively, the scalar $\eta$ can develop a vacuum expectation value for contributing to the dark photon mass. However, the dark photon mass could not be ultralight in this case. This is because the scalar $\eta$ usually can not spontaneously develop an extremely small vacuum expectation value, meanwhile, the gauge coupling $g_X^{}$ usually can not naturally acquire a very tiny value as shown in Eq. (\ref{gx}).

\section{Conclusion}

In this paper we have demonstrated that the artificially introduced $U(1)_X^{}$ gauge group for dark photon and the SM $U(1)_Y^{}$ gauge group for hypercharge can simultaneously originate from two $U(1)_1^{}\times U(1)_2^{}$ gauge groups under which the SM Higgs scalar and chiral fermions carry the same $U(1)_1^{}$ and $U(1)_2^{}$ charges without causing any gauge anomalies. In this scenario, the $U(1)_X^{}\times U(1)_Y^{}$ kinetic mixing can not be induced at all if the $U(1)_1^{}$ gauge coupling is exactly equal to the $U(1)_2^{}$ gauge coupling. By further introducing a spontaneously broken mirror symmetry between the $U(1)_1^{}$ and $U(1)_2^{}$ gauge groups, we can naturally obtain a one-loop induced difference between the $U(1)_1^{}$ and $U(1)_2^{}$ gauge couplings. In consequence, the $U_X^{} \times U_Y^{}$ kinetic mixing can be highly suppressed in a natural way.

\textbf{Acknowledgement}: This work was supported in part by the National Natural Science Foundation of China under Grant No. 12175038.

\end{document}